\documentclass[%
 reprint,
 amsmath,amssymb,
 aps,
]{revtex4-2}

\usepackage{graphicx}
\usepackage{dcolumn}
\usepackage{bm}

\begin{document}

\preprint{APS/123-QED}

\title{Quantum State Preparation via Nested Entanglement} 

\author{Geoffrey L. Warner}
\affiliation{%
 The MITRE Corporation, Charlottesville, VA 22911 USA
}%




\date{\today}

\begin{abstract}
We develop a representation of an $n$-qubit register that parameterizes its statevector as a series of nested entanglements.  We show that the recursive substructure of this representation provides a natural framework for automating the construction of quantum circuits for state preparation.  It also allows for a straightforward treatment of pure state separability.  We discuss a novel derivation of uniformly controlled rotations and the quantum Fourier transform within this representation, and consider the effects of single-qubit basis changes on its overall structure.  We end with a discussion of the apparent connection between the compressibility of the state description in this representation, and the circuit complexity required to prepare it.   
\end{abstract}

\maketitle


\section{\label{sec:intro}Introduction}

A quantum algorithm is a procedure for manipulating the qubits of a quantum register into a collective state encoding the solution to some mathematical problem.  Interest in such algorithms stems from the widespread belief that there are classes of problem for which the computational burden imposed by ordinary methods of solution may be significantly reduced through the exploitation of entanglement \cite{Vidal2003, Jozsa2003}.  Comparisons of the relative burdens imposed by different algorithms are usually made in terms of the asymptotic scaling of computational resources such as circuit size or depth.  

A necessary part of many quantum algorithms is the preparation of arbitrary entangled states from an initial separable state of the register.  In the quantum circuit model of quantum computation, the state preparation problem can be described in terms of a sequence of operations whose combined action produces the desired state.  Each step in this sequence is composed of multiple elementary gates acting simultaneously on each qubit.   These elementary gates may act independently upon their target qubits, or else collectively in a manner that depends irreducibly on the states of multiple qubits; in the latter case, the corresponding operators are sometimes called entangling gates.  

The state preparation problem consists in specifying the possible structures of quantum circuits capable of producing the desired target state, ideally with good asymptotic scaling for some measure of computational complexity.  This is, in general, a difficult problem, and will depend upon certain details of the hardware governing qubit control, including the mechanisms available for performing single qubit rotations, and the physical implementation of inter-qubit coupling.   While such details are important, it is often advantageous to consider the problem more abstractly in terms of the different possible circuits that can be composed from an elementary gate set.  Multiple efforts in this direction \cite{Plesch2011, Znidaric2008, Kaye2001, Goubault2020, Mottonen2004, Mottonen2005, Rakyta2022} have yielded circuit construction schemes whose scaling with respect to certain measures of circuit complexity, such as size, depth, or the number of entangling gates, is close to optimal.  

In this paper, we develop a hierarchical representation of an arbitrary quantum register in terms of the nested entanglement of each qubit with its successors in an ordered list.  Such a representation affords a unified framework for parameterizing the entanglement structure of an arbitrary target state.  Further, as this representation corresponds to a simple binary tree, it allows for the application of conventional computational methods to the automation of quantum circuits for state preparation.  In particular, it allows for the exploitation of recursive substructure, and makes evident how reductions in circuit complexity can arise from certain patterns among its parameters.

We begin by describing a procedure for computing the parameters of the nested representation via postorder traversal of the tree.  We then embark on a discussion of two decomposition schemes for converting a given tree structure into a quantum circuit.  The first of these is a subtree decomposition that we use to derive a general recursion relation for the entangling part of a state preparation circuit.  This recursion relation is then used to establish a simple criterion for pure state separability of the tree.

A second, ``pyramidal" decomposition scheme is then explored, and shown to be equivalent to the formalism of uniformly controlled rotations.  We then apply the pyramidal decomposition to a derivation of the quantum Fourier transform, and discuss how the parametric structure of its tree representation leads to the polynomial scaling of the transform itself.  Finally, we proceed to a discussion of repeated or redundant substructure within the tree formalism, and its removal by single-qubit changes of basis.  We also provide a different point of view on the generalized Schmidt decomposition derived by Carteret et. al. \cite{Carteret2000} by casting it as a complex multilinear system.  We end with a discussion of the apparent connection between the compressibility of a given state's tree representation, and the complexity of the circuit required to produce it.  

\section{Nested Entanglement Trees}
Consider an $n$-qubit register.  We define the separable state $\lvert b_{0}^{n} \rangle = \lvert \uparrow \uparrow \cdots \uparrow \rangle \equiv \lvert 0 \rangle $ as the fixed reference from which all other possible states may be generated by unitary transformations, and assign to each qubit in the register an ordinal index $0, 1, \dots, n-1$ proceeding from left to right.  Following the usual convention, we make the identifications $\lvert \uparrow \rangle_{i} = \lvert 0 \rangle_{i}$ and $\lvert \downarrow \rangle_{i} = \lvert 1 \rangle_{i}$ for qubit $i$.  Any arbitrary target state $\lvert \Omega \rangle$  of the register can be written as a superposition over the $N = 2^{n}$ basis states $\lvert b_{k}^{n} \rangle$, where $b_{k}^{n}$ is the binary representation of $k$ with $n$ digits, as 
\begin{equation}\label{final_state}
\lvert \Omega \rangle = \sum_{k=0}^{N-1} \gamma_{k} | b_{k}^{n} \rangle.
\end{equation}
Here the $\gamma_{k}$ are complex numbers, each with 2 degrees of freedom.  Accounting for overall normalization and a global phase, this leaves $2N - 2$ total parameters to be specified. 

In this paper, we consider a representation of a general $n$-qubit register that treats the wavefunction as a sequence of nested entanglements of the form 
\begin{equation}\label{recursive_psi}
\lvert \Psi_{n - k} \rangle = \alpha_{k} \lvert \uparrow \rangle_{k} \otimes \lvert \Psi_{n - k-1} \rangle  +  \beta_{k}  \lvert \downarrow \rangle_{k} \otimes \lvert \Psi_{n - k -1}' \rangle,
\end{equation}
where $\alpha_{k}, \beta_{k}$ are complex coefficients satisfying $|\alpha_{k}|^{2} + |\beta_{k}|^{2} = 1$; $\vert \Psi_{n - k} \rangle$ is an arbitrary state defined on the $n - k$ qubits labeled $k, k+1, \dots, n-1$; $\lvert \uparrow \rangle_{k}, \lvert\downarrow \rangle_{k}$ are the two basis states of the $k$-th qubit; and $| \Psi_{n - k -1} \rangle$, $| \Psi_{n - k -1}' \rangle$ are two arbitrary (normalized) states defined on the $n - k -1$ qubits labeled $k+1, k+2, \dots, n-1$.  Because these latter wavefunctions represent arbitrary states of $m = n-k-1$ qubits, we know that each requires $2^{m} - 2$ parameters to specify.  The $\alpha_{k}, \beta_{k}$ coefficients add only 2 parameters to this, so that the total number of parameters in $|\Psi_{n - k} \rangle $ is $2 \cdot (2^{m} - 2) + 2 = 2^{m+1} - 2$, as expected.  By induction, we see that we can generate the whole hierarchy of multi-qubit states recursively, provided we specify that at the lowest level, all single qubit states are of the form $\lvert \Psi_{1} \rangle = \alpha_{n-1} \lvert \uparrow \rangle + \beta_{n-1} \lvert \downarrow \rangle$. 

\begin{figure}[hbt]
 \begin{center}
  \includegraphics[height=1.90 in]{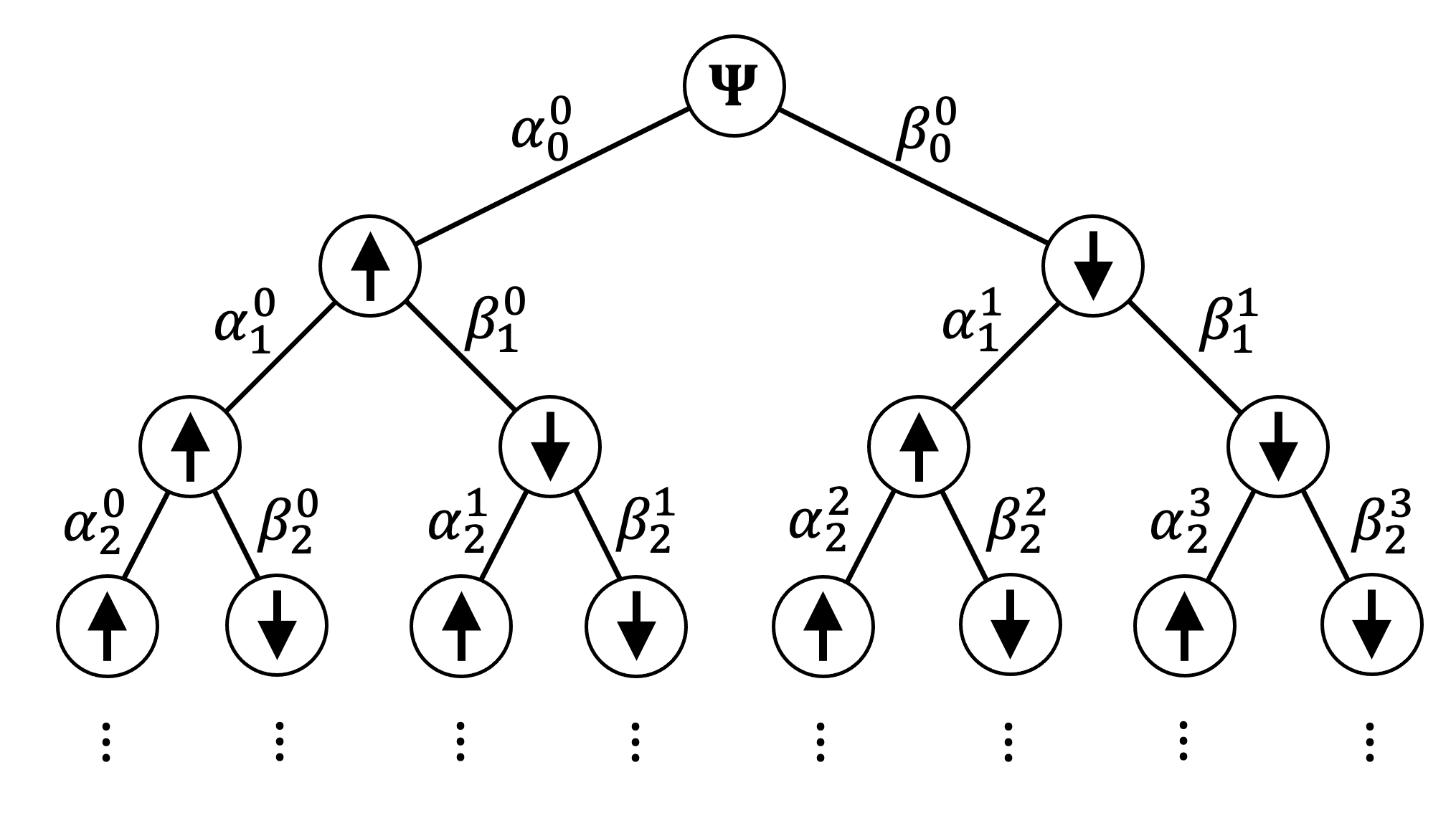}
 \caption{Nested entanglement representation.  }\label{fig:nested_representation}
 \end{center}
\end{figure}

The recursive structure of $\Psi_{n}$ can be represented by a binary tree with $2^{n} - 1$ interior nodes, as depicted in Figure \ref{fig:nested_representation}.  For convenience, let us call the associated data structure, which stores the values of the coefficients $\alpha^{i}_{j}, \beta^{i}_{j}$  at the corresponding nodes, the ``$\psi$-tree."  Here the subscript indicates the depth of the node, starting from $0$ at the root node, while the superscript labels its position within a given level, starting from $0$ at the leftmost node.  Without loss of generality, we may choose a symmetrical parameterization of these coefficients in terms of Bloch sphere angles $(\theta, \phi)$  so that $\alpha^{i}_{j} = \cos (\frac{1}{2}\theta^{i}_{j}) e^{-\frac{i}{2}\phi^{i}_{j}}$ and $\beta^{i}_{j} = \sin (\frac{1}{2}\theta^{i}_{j}) e^{\frac{i}{2}\phi^{i}_{j}}$.

In addition to the picture described above, we can represent the state corresponding to the $\psi$-tree as a product of unitary operators whose action on $\lvert b^{n}_{0} \rangle$ produces the nested superpositions depicted in Figure \ref{fig:nested_representation}.  Let us define a set of single qubit unitaries $U^{i}_{j} \in SU(2)$ as $U^{i}_{j} = \begin{pmatrix} \alpha^{i}_{j} & -\beta^{i*}_{j} \\ \beta^{i}_{j} & \alpha^{i*}_{j} \end{pmatrix}$.  For this choice of $U$, the  symmetrical parameterization of the $\alpha, \beta$ coefficients corresponds to the simple unitary matrix $ U^{i}_{j} = R_{z}(\phi^{i}_{j})R_{y}(\theta^{i}_{j})$, where $R_{z}(\phi) = \begin{pmatrix} e^{-i \frac{\phi}{2}} & 0 \\ 0 & e^{i\frac{\phi}{2}} \end{pmatrix}$ and $R_{y}(\theta)  = \begin{pmatrix} \cos \frac{1}{2} \theta & -\sin \frac{1}{2} \theta \\ \sin \frac{1}{2} \theta  & \cos \frac{1}{2} \theta  \end{pmatrix}$.  The restriction to $SU(2)$ greatly simplifies the resulting formalism and allows for a straightforward treatment of entanglement, as will be seen below.

In the construction of a given $\psi$-tree, it may sometimes occur that an interior node has both $\alpha, \beta = 0$.  In such cases, the corresponding unitary operator is undefined.  One may assign any arbitrary unitary to these nodes without affecting the resulting wavefunction.  We will call these \textit{dead nodes} to distinguish them from nodes with a determinate unitary.  They inevitably occur as daughters of nodes where $R_{y} = R_{y}(0)$ or $R_{y}(\pi)$, in which case they form the root node of a dead subtree.  

Before we proceed, we must define some useful notation.  First, we denote by $P^{k}_{\uparrow}$, $P^{k}_{\downarrow}$ the projection operators $\lvert \uparrow \rangle \langle \uparrow \rvert$, $\lvert \downarrow \rangle \langle \downarrow \rvert$ acting on the $k$-th qubit, and write $C^{k}(U) = P^{k}_{\uparrow} + P^{k}_{\downarrow}U$ for the controlled-$U$ operator.  In cases where we wish to distinguish the state of the control qubit, we may sometimes use the redundant notation $C^{k}_{\downarrow}, C^{k}_{\uparrow}$.  Second, we write the adjoint operation as $U^{\dagger} = \widetilde{U}$ in order to avoid notational difficulties associated with the many subscripts and superscripts attached to the various operators.  Finally, we define $T^{i}_{j}$ as the matrix product that generates a particular superposition on the subtree whose root node is at $(i,j)$.  This implies, by definition, that $|\Psi_{n} \rangle = T^{0}_{0} |0 \rangle$, and  that
\begin{equation}\label{subtree_recursion}
T^{i}_{j} = \left( P^{j}_{\uparrow} T^{2i}_{j+1} + P^{j}_{\downarrow} T^{2i+1}_{j+1} \right) U^{i}_{j}.
\end{equation}
This equation captures the recursive substructure of the $\psi$-tree.   Each branch point or interior node of this tree forms the root node of a subtree whose statevector is determined by the product of two operators:  first, a single qubit unitary acting on the root node qubit, and second, an operator comprising two subtrees that each rotate the remaining qubits in a manner that depends on the state of the root node.

In the following two sections, we consider two distinct operator decomposition schemes within the nested entanglement picture.  The first of these is a subtree decomposition that breaks the state preparation problem down to the level of multiply controlled single qubit unitaries.  The resulting operator product provides a simple and intuitive criterion for separability, and may be useful in situations where multiply-controlled or multi-target unitaries can be treated as elementary gates.  The second, or ``pyramidal" scheme, constructs the operator product level by level within the tree.  The resulting product can be shown to be related to the formalism of uniformly controlled rotations developed in \cite{Mottonen2004} .  By reversing the order of the operators corresponding to each level in a pyramidal decomposition, we show that the resulting gate sequence corresponds to a basis transformation of the computational states.  We then derive the quantum Fourier transformation within this formalism. 

\subsection{Computation of Tree Parameters}
In order to proceed, we must first devise a method for converting an arbitrary target state (\ref{final_state}) of the register into its corresponding tree.  The $\psi$-tree for an arbitrary $n$-qubit state provides a kind of blueprint for constructing that state in terms of a set of parameters $\{\theta^{i}_{j}, \phi^{i}_{j} \}$ defined at each interior node of the tree.   The overlap  $\chi = \langle \Omega | \Psi(\{\theta^{i}_{j}, \phi^{i}_{j} \}) \rangle$ between the target state $\lvert \Omega \rangle$ and the state $\lvert \Psi(\{ \theta^{i}_{j}, \phi^{i}_{j} \}) \rangle$ determined by these parameters defines an objective function  $\Phi = |\chi|^{2}$.  The parameters required to generate $\lvert \Omega \rangle$ correspond to those that obtain when $\Phi = 1$, which is the maximum of the objective function.  They can be discovered by varying $\{\theta^{i}_{j}, \phi^{i}_{j} \}$ until $\Phi$ reaches its maximum, as by gradient ascent, or else by a more direct procedure, which we describe below.

First, let us define a secondary data structure called the ``$\chi$-tree" whose form precisely mirrors that of the $\psi$-tree.  The root node of the $\chi$-tree contains the value $\chi$, whereas the leaf nodes $\chi_{n}^{i}$ are assigned the values $\chi_{n}^{i} = \gamma_{i}$ appearing in the target state $\lvert \Omega \rangle$ of Equation (\ref{final_state}).  Propagating upwards from the leaf nodes, we can compute the $\alpha_{j}^{i}$, $\beta_{j}^{i}$ coefficients at each interior node of the $\psi$-tree by accumulating values up the $\chi$-tree according to the formula   
\begin{equation}
\chi^{i}_{j} = \alpha^{i}_{j} \chi^{2i}_{j+1} + \beta^{i}_{j} \chi^{2i+1}_{j+1}
\end{equation}
where $\chi^{i}_{j} = |\chi^{i}_{j}|e^{i \xi^{i}_{j}}$ and the $\alpha, \beta$ values are chosen as
\begin{eqnarray}
\alpha^{i}_{j} &=& \frac{|\chi^{2i}_{j+1}|}{\sqrt{|\chi^{2i}_{j+1}|^{2} + |\chi^{2i+1}_{j+1}|^{2}}} e^{i(\xi^{2i}_{j+1} - \xi^{2i+1}_{j+1})/2}\nonumber \\ 
\beta^{i}_{j} &=& \frac{|\chi^{2i+1}_{j+1}|}{\sqrt{|\chi^{2i}_{j+1}|^{2} + |\chi^{2i+1}_{j+1}|^{2}}} e^{-i(\xi^{2i}_{j+1} - \xi^{2i+1}_{j+1})/2},
\end{eqnarray}
a form which preserves the symmetrical parameterization described above.  This yields
\begin{equation}
\chi^{i}_{j} = \sqrt{|\chi^{2i}_{j+1}|^{2} + |\chi^{2i+1}_{j+1}|^{2}}e^{i(\xi^{2i}_{j+1} + \xi^{2i+1}_{j+1})/2}
\end{equation}
for each parent node.
This calculation can be effected by means of a simultaneous postorder traversal of the $\psi$- and $\chi$-trees, which is ${\cal{O}}(N)$.
The resulting set of relations completely determines all coefficients up the tree, ending at the root node, which takes the value
\begin{equation}
\chi^{0}_{0} = \chi = \sqrt{\sum^{N-1}_{k=0} |\gamma_{k}|^{2}}e^{i\xi} = e^{i\xi}
\end{equation}
so that $\Phi = |\chi|^{2} = 1$.

Conversely, one may compute the coefficients of the wavefunction in the computational basis from its $\psi$-tree representation by multiplying the amplitudes down the corresponding branches or paths in the tree.

\section{Subtree decomposition}

We begin by rewriting Equation (\ref{subtree_recursion}) in the form
\begin{equation}\label{recursion}
T^{i}_{j} = C^{j}(T^{2i+1}_{j+1} \widetilde{T}_{j+1}^{2i}) T^{2i}_{j+1} U_{j}^{i}.
\end{equation}
This relation, in combination with the identities
\begin{eqnarray}\label{identities}
C^{k}(AB) &=& C^{k}(A)C^{k}(B) \\
C^{k}(U^{-1}) &=& C^{k}(U)^{-1} \\
(AB)^{-1} &=& B^{-1} A^{-1} \\
C^{k}(U) U^{-1} &=& X_{k} C^{k}(U^{-1}) X_{k}
\end{eqnarray}
can be used to reduce any $\psi$-tree operator $T_{0}^{0}$ to a product of multiply-controlled, single-qubit unitaries acting on each qubit in the register.  The precise form of the decomposition will depend upon the order of application of these identities.  The resulting sequence can be broken down into more elementary gates by applying the identities (\ref{first_identity})-(\ref{last_identity}) appearing below in Section \ref{pyramidal_section}.

Here we specify a particular decomposition structure by interpreting Equation (\ref{recursion}) in a slightly different way.  The full tree for $n$ qubits is composed of two subtrees defining the state of $n-1$ qubits.  If we know the form of the general solution for $n-1$, we can find the solution for $n$ by shifting the register to the right by one qubit, adding a qubit in the ``0" position, and substituting the shifted solutions into Equation (\ref{recursion}).  This results in a recursion of the form
\begin{equation}\label{recursive_2}
T^{0}_{0,n} = C^{0}(T^{1}_{1, n-1}) C^{0}(\widetilde{T}^{0}_{1, n-1}) T^{0}_{1, n-1} U_{0}^{0}
\end{equation}
where $T_{0,n}^{0}$ is the operator generating an arbitrary $n$-qubit state, and $T_{1, n-1}^{0}$, $T^{1}_{1, n-1}$ each generate arbitrary states on the remaining $n-1$ qubits.  As an illustration, we note that, since $T_{0,1}^{0} = U_{0}^{0}$, two applications of Equation (\ref{recursive_2}) yields
\begin{align}
T_{0,2}^{0} &=  C^{0}(U^{1}_{1}\widetilde{U}^{0}_{1}) U^{0}_{1} U^{0}_{0} \nonumber \\
T_{0,3}^{0} &= C^{0,1}(U_{2}^{3} \widetilde{U}_{2}^{2}) C^{0}(U_{2}^{2} \widetilde{U}_{2}^{0}) C^{0}(U_{1}^{1} \widetilde{U}_{1}^{0} ) \nonumber \\
&\times X_{0} C^{0,1}(U_{2}^{1} \widetilde{U}_{2}^{0})) X_{0} U_{2}^{0} U_{1}^{0} U_{0}^{0}
\end{align}
where we have followed the rule that  $T^{i}_{j} = U^{i}_{j}$ whenever $j = n-1$.  We have also introduced the notation  $C^{i, j}() = C^{i}(C^{j}())$ for convenience.

Note that the two and three qubit solutions have the form of a product of single qubit unitaries corresponding to the leftmost interior nodes in each level, followed by a product of entangling operators each comprising at least one controlled unitary.  We write the $(n-1)$-qubit solutions of this form for the right and left subtrees as
\begin{eqnarray}
 T_{1,n-1}^{0} &=& \Gamma_{1,n-1}^{0} \prod_{k=1}^{n-1} U_{k}^{0} \nonumber \\
T_{1,n-1}^{1} &=& \Gamma_{1,n-1}^{1}\prod_{k=1}^{n-1} U_{k}^{2^{k-1}}.
\end{eqnarray}  
Here $\Gamma$ denotes the product of entangling operators described above.  Substitution of these expressions into (\ref{recursive_2}) yields
\begin{equation}
T_{0,n}^{0} = \Gamma_{0,n}^{0} \prod_{k=0}^{n-1} U_{k}^{0},
\end{equation}
where 
\begin{equation}\label{entanglement_recursion}
\Gamma_{0,n}^{0} 
= C^{0}_{\downarrow}(\Gamma_{1,n-1}^{1}) C^{0}_{\uparrow}(\Gamma_{1,n-1}^{0}) \left[ \prod_{k=1}^{n-1} C^{0}_{\downarrow}(U_{k}^{2^{k-1}} \widetilde{U}_{k}^{0}) \right] ,
\end{equation}   
 which preserves the basic form.  Hence, by induction on (\ref{recursive_2}), we see that this same form must obtain for all $n>3$ as well.  Moreover, it is clear that each operator in $\Gamma$ must have the general form $C^{0,1, \dots, k}(U_{i}^{a} \widetilde{U}_{i}^{b})$ (or its similarity transformations under $X_{j}$ for $0 \leq j \leq k$),  where $U_{i}^{a} \widetilde{U}_{i}^{b} = R_{z}(\phi_{i}^{a}) R_{y}(\theta_{i}^{a} - \theta_{i}^{b}) R_{z}(-\phi_{i}^{b})$ are single-qubit operators parameterized by the Euler angles $(\phi_{i}^{a}, \theta_{i}^{a} - \theta_{i}^{b}, -\phi_{i}^{b} )$.  This is evident by induction on (\ref{entanglement_recursion}) starting from the solution for 2 qubits.  

\subsection{Separability}
A distinct advantage of the nested entanglement picture is the simplicity of the corresponding criterion for separability.  Pure state separability is a topic of perennial interest \cite{Jorrand2003, Yu2006, Makela2010, Neven2018}.  A pure state is separable if it factors into a product of single qubit states.  Since by construction the $\Gamma$ operators defined above contain all of the entangling gates in $T_{0,n}^{0}$, it seems reasonable to suppose that the state it generates will be separable if and only if $\Gamma_{0,n}^{0} = 1$.   The sufficiency of this statement follows trivially by definition, but its necessity does not.  In order to demonstrate necessity, we shall find it expedient to first condition the $\psi$-tree so that it assumes a certain canonical form, which we define below.  

A particular $\psi$-tree may be brought into canonical form by a simple procedure.  First, we note that any valid tree must, by definition, contain at least one path of nonzero amplitude spanning its full depth.  If the leftmost branch of the tree corresponds to such a path, then the first step is already accomplished.  If not, we must find another such path in the tree.  This path can then be transformed into the leftmost branch by applying the similarity transformations 
\begin{eqnarray}
\lvert b^{n}_{0} \rangle &\rightarrow& X_{j} \lvert b^{n}_{0} \rangle \nonumber \\
T_{0,n}^{0} &\rightarrow& X_{j} T_{0,n}^{0} X_{j}
\end{eqnarray}
on those levels $j$ where the path veers rightward.  These transformations define equivalence classes of $\psi$-trees that map to the same wavefunction under redefinitions of ``up" and ``down" for a given qubit.  Note that the effect of the second transformation will be to induce the transformations
\begin{eqnarray}
P^{j}_{\uparrow}, P^{j}_{\downarrow} &\rightarrow& P^{j}_{\downarrow}, P^{j}_{\uparrow} \nonumber \\
U_{j}^{i} =  R_{z}(\phi_{j}^{i}) R_{y}(\theta_{j}^{i}) &\rightarrow&  R_{z}(-\phi_{j}^{i}) R_{y}(-\theta_{j}^{i}).
\end{eqnarray}

The second step consists in assigning to each dead node a unitary matrix equal to that of the leftmost node at the corresponding level.  In adopting this convention, we exploit the freedom noted above to choose dead node values according to convenience.

Imagine that we have brought a given $\psi$-tree into canonical form, and that we have constructed the corresponding $T^{0}_{0,n}$ from this tree.  As above, this operator splits into an entangling part $\Gamma = \Gamma^{0}_{0,n}$, and a product of single qubit unitaries, $U = \prod_{k=0}^{n-1} U_{k}^{0}$.  Let us call the separable part of the state $\lvert \phi \rangle = U \lvert b^{n}_{0} \rangle$.  If $\Gamma \neq 1$, and $U$ consists only of non-diagonal unitaries, then $\Gamma \vert \phi \rangle \neq \lvert \phi \rangle$, and $\Gamma = 1$ is a necessary condition for separability.  However, it may happen that some of these unitaries are diagonal, in which case it is possible that $\Gamma$ acts trivially on $\lvert \phi \rangle$, i.e. $\Gamma \lvert \phi \rangle = \lvert \phi \rangle$, even if $\Gamma \neq 1$.  We now show that this cannot happen if the tree is in canonical form.

To begin, we must first prove that $\Gamma = 1$ if and only if the $U^{i}_{j}$ satisfy a certain condition level by level in the tree.  Rewriting the right hand side of  (\ref{entanglement_recursion}), we see that $\Gamma = 1$ implies 
\begin{equation}
P^{0}_{\uparrow} \Gamma_{1,n-1}^{0} + P^{0}_{\downarrow} \Gamma_{1,n-1}^{1} \left[ \prod_{k=1}^{n-1} U_{k}^{2^{k-1}} \widetilde{U}_{k}^{0} \right] = 1.
\end{equation}
This can only hold if the factors multiplying each projector separately equate to unity:  
\begin{eqnarray}
\Gamma_{1,n-1}^{0} &=& 1 \nonumber \\
\Gamma_{1,n-1}^{1} \left[ \prod_{k=1}^{n-1} U_{k}^{2^{k-1}} \widetilde{U}_{k}^{0} \right] &=& 1.
\end{eqnarray}
Because $\Gamma_{1,n-1}^{1}$ comprises only entangling gates, the second equality can hold only if two conditions are met; namely, a) $U_{k}^{2^{k-1}} = U_{k}^{0}$ for all $k$ in the product, and b)  $\Gamma_{1,n-1}^{1} = 1$.  But this only repeats the same argument for the remaining $n-1$ qubits in each of the two subtrees.  Hence, we conclude that, if $\Gamma = 1$, then
\begin{equation}\label{unitary_condition}
U^{i}_{j} = U^{0}_{j} \textrm{   for all } i,j .
\end{equation}
That the converse is true can be seen by noting that the argument can be run the other way, starting from single qubit states at level $n-1$, for which $\Gamma = 1$ by definition, and proceeding up the hierarchy by applying (\ref{unitary_condition}) at each stage.

To show that $\Gamma = 1$ is a necessary condition for separability, we consider the case where (\ref{unitary_condition}) is violated by a single node at $(j,k)$, $j \neq 0$.  It is clear that $(j,k)$ cannot lie on a dead branch of the tree, since this would contradict the construction of the canonical form, which assigns to each dead node the value of the leftmost unitary on the same level.  Hence we see, by Eq. (\ref{entanglement_recursion}), that $\Gamma$ must contain an operator of the form $U^{j}_{k} \tilde{U}^{0}_{k}$ that is nested in a chain of control with a nonzero amplitude in the register wavefunction; the action of such an operator on $\lvert \phi \rangle$ is manifestly nontrivial.   This implies that the condition $\Gamma = 1$, and hence (\ref{unitary_condition}), is both necessary and sufficient for separability when the $\psi$-tree is in canonical form.

Thus, to assess whether a given pure state is separable, it suffices to construct its $\psi$-tree level by level until one observes a violation of (\ref{unitary_condition}).  Of course, in general this can be an exponential proposition, since for an arbitrary state one must visit $2^{n} - 1$ nodes to construct its tree; however, in cases where $\lvert \Omega \rangle$ has certain regularities, or sets of adjacent zeroes, the problem may be greatly simplified.  Note that the present formalism cannot be applied to the problem of mixed state separability, for which a more general construction involving, e.g., concurrences or entanglement witnesses, is required \cite{Horodecki2009}.

\subsection{Extension of the Bloch sphere concept to multiple qubits}
Since each node $(i,j)$ of the $\psi$-tree contains two angle parameters $\theta^{i}_{j}$ and $\phi^{i}_{j}$, it can be taken to provide a formal generalization of the Bloch sphere construction to multiple qubits.  The construction consists in assigning a separate Bloch sphere to each and every node of the tree.  Such a construction may be useful for visualizing the states of few qubit systems, or states exhibiting certain regularities among their coefficients.  

\section{Pyramidal decomposition}\label{pyramidal_section}

In addition to the subtree decomposition described above, one may regard any arbitrary state of the quantum register as comprising a product of operators corresponding to each level of the full tree.  That is, the state may be written as a product over levels $k$ of the form $\prod_{k=n-1}^{0} \Lambda^{0}_{0,k} \lvert 0 \rangle$ where 
\begin{equation}\label{pyramidal_recursion}
\Lambda^{i}_{j,k} = P^{j}_{\uparrow} \Lambda^{2i}_{j+1, k} + P^{j}_{\downarrow} \Lambda^{2i+1}_{j+1,k}
\end{equation}
and
\begin{equation}
\Lambda^{i}_{k-1,k} = P^{k-1}_{\uparrow} U^{2i}_{k} + P^{k-1}_{\downarrow} U^{2i+1}_{k}.
\end{equation}
Each level may be regarded as a nested sequence of operators containing, at its deepest stage of recursion, all the $2^{k}$ unitaries acting on qubit $k$.  More explicitly, the sequence of operators takes the form
\begin{eqnarray}
\Lambda_{0,0}^{0} &=& U_{0}^{0}\nonumber \\
\Lambda_{0,1}^{0} &=& P^{0}_{\uparrow} U_{1}^{0} + P^{0}_{\downarrow} U_{1}^{1} \nonumber \\
\Lambda_{0,2}^{0} &=& P^{0}_{\uparrow} (P^{1}_{\uparrow} U_{2}^{0} + P^{1}_{\downarrow} U_{2}^{1}) + P^{0}_{\downarrow} (P^{1}_{\uparrow} U_{2}^{2} + P^{1}_{\downarrow} U_{2}^{3}) \nonumber \\
&\vdots& 
\end{eqnarray}

Recall that for the symmetrical parameterization, all single qubit unitaries $U^{i}_{j}$ take the form $R_{z}(\phi^{i}_{j})R_{y}(\theta^{i}_{j})$, so that $\Lambda_{0,0}^{0} = R_{z}(\phi^{0}_{0})R_{y}(\theta^{0}_{0})$.  As we shall see, all subsequent levels can be reduced to products of $y$- and $z$-rotations and CNOT gates by making use of the following elementary identities:
\begin{align}
X_{k}^{2} &= 1 \label{first_identity} \\
R_{y,z}(\alpha_{k} + \beta_{k}) &= R_{y,z}(\alpha_{k}) R_{y,z}(\beta_{k}) \\
R_{y,z}(\alpha_{k}) X_{k} &= X_{k} R_{y,z}(-\alpha_{k}) \label{x_identity} \\
\label{factoring}
P^{j}_{\uparrow} AB + P^{j}_{\downarrow} CD &= \left(P^{j}_{\uparrow} A + P^{j}_{\downarrow} C \right) \left(P^{j}_{\uparrow} B + P^{j}_{\downarrow} D \right) \\
X_{k} C^{j}_{k} X_{k} &= C^{j}_{k} \label{last_identity}
\end{align}
Here we have defined the quantity $C^{i}_{j} = P^{i}_{\uparrow} + P^{i}_{\downarrow} X_{j}$, which is nothing but a CNOT gate where the control qubit is $i$, and the target qubit is $j$.

The basic strategy of this approach is to factorize each $\Lambda^{0}_{0,k}$ into a ``$z$-block" and a ``$y$-block" of operators, where each block comprises products of their respective rotations interleaved with CNOT gates.  As the $k=0$ operator is already factorized, we illustrate this strategy by considering the $k=1$ case below:
\begin{align}
\Lambda^{0}_{0,1} &=P^{0}_{\uparrow} R_{z}(\phi_{1}^{0})R_{y}(\theta_{1}^{0} )+ P^{0}_{\downarrow} R_{z}(\phi_{1}^{1})R_{y}(\theta_{1}^{1}) \nonumber \\
&=  \left(P^{0}_{\uparrow} R_{z}(\phi_{1}^{0})+ P^{0}_{\downarrow} R_{z}(\phi_{1}^{1}) X_{1}) \right) \nonumber \\
&\times \left(P^{0}_{\uparrow} R_{y}(\theta_{1}^{0} )+ P^{0}_{\downarrow} X_{1}R_{y}(\theta_{1}^{1} ) \right). \nonumber \\
\end{align}
Factoring out an $R_{z}(\phi_{1}^{0}) $ on the left and an $ R_{y}(\theta_{1}^{0} )$ on the right, and applying identity (\ref{x_identity}) above, yields
\begin{align}
\Lambda^{0}_{0,1} &= R_{z}(\frac{\phi_{1}^{0} + \phi_{1}^{1}}{2} ) C^{0}_{1}  R_{z}(\frac{\phi_{1}^{0} - \phi_{1}^{1}}{2} ) \nonumber \\
&\times R_{y}(\frac{\theta_{1}^{0} - \theta_{1}^{1}}{2} ) C^{0}_{1}  R_{y}(\frac{\theta_{1}^{0} + \theta_{1}^{1}}{2})
 \end{align}
One can proceed to calculate all $\Lambda^{0}_{0,k}$ by this same procedure; that is, by substituting the factorization for level $k-1$  into level $k$, with appropriate transformations of the indices, via (\ref{pyramidal_recursion}).   This leads to a factorization at level $k$ that doubles the number of rotation operators at $k-1$, and introduces some new CNOT gates between them.  From this we conclude that the number of $R_{z}$ (and $R_{y}$) gates in level $k$ is $r_{k} = 2^{k}$, and hence that the number of CNOT gates is $c_{k} = 2(r_{k} - 1)$.  The total number of CNOT gates for $n$ qubits is therefore $\sum_{l=1}^{n} c_{k}  = 2^{n+1} - 2n - 2$, which is precisely the scaling achieved by the use of uniformly controlled rotations \cite{Mottonen2004, Mottonen2005}.
 
Note that the $R_{z}$ operators can be pulled left through the $R_{y}$ operators belonging to lower levels, so that the two sets of operators can be consolidated into separate $z$ and $y$ blocks.  The resulting decomposition indeed bears a striking resemblance to a sequence of uniformly-controlled rotations.  They are, in fact, equivalent, as can be seen by the following argument.  
 
 Let us denote by $C^{b_{k}^{n}}(U)$ the operator corresponding to the nested control sequence specified by the binary number $b_{k}^{n}$, where the value of the $i$-th digit indicates the state of the $i$-th control qubit.  It is clear that $b_{k}^{n}$ describes a path through $n$ levels of the tree, provided we take ``0" to indicate a leftward turn, and ``1" a rightward one, starting from the root node.  If the operator is nested within higher levels of control, then the associated path begins at the terminal node of that higher level sequence, so that 
 \begin{equation}
 C^{b^{n}_{k}}(U) = C^{b^{l}_{i}}(C^{b^{n-l}_{j}}(U))
 \end{equation}
 where $k = 2^{l}i + j$.  

Any two such operators whose root nodes coincide but whose terminal nodes lie on divergent branches must always commute.  This can be seen by noting that,  for operators $C^{b^{n}_{k}}(U)$, $C^{b ^{n'}_{k'}}(V)$, if $b^{n}_{k}$, $b ^{n'}_{k'}$ diverge at level $l$, so that $C^{b^{n}_{k}}(U) = C^{b^{l-1}_{m}}(P^{l}_{\uparrow} + P^{l}_{\downarrow} C^{b^{n-l}_{i}}(U))$ and $C^{b^{n'}_{k'}}(V) = C^{b^{l-1}_{m}}(P^{l}_{\uparrow}C^{b^{n'-l}_{j}}(V) + P^{l}_{\downarrow} )$, then
\begin{eqnarray}
C^{b^{n}_{k}}(U) \cdot C^{b ^{n'}_{k'}}(V) &=& C^{b^{l-1}_{m}}(P^{l}_{\uparrow}C^{b^{n'-l}_{j}}(V) + P^{l}_{\downarrow} C^{b^{n-l}_{i}}(U)) \nonumber \\
&=& C^{b ^{n'}_{k'}}(V) \cdot C^{b^{n}_{k}}(U)
\end{eqnarray}
 
 It is not difficult to see that, by repeated application of identity (\ref{factoring}), the level operators can be written
\begin{align}
\Lambda^{0}_{0,l} &= \prod_{k=0}^{2^{l}-1} C^{b^{l}_{k}}(U^{k}_{l}) \\
&=  \prod_{k=0}^{2^{l}-1} C^{b^{l}_{k}}(R_{z}(\phi^{k}_{l}))  \prod_{k=0}^{2^{l}-1} C^{b^{l}_{k}}(R_{y}(\theta^{k}_{l})) 
\end{align}
where the second step follows from (\ref{factoring}) and from the commutation rules described above.  But each block of $z$ and $y$ rotations above is equivalent to a uniformly controlled rotation as defined in \cite{Mottonen2004} .  Further, because the $R_{z}$ operators are diagonal they commute with the projectors $P^{k}_{\uparrow}, P^{k}_{\downarrow}$, so that  $\left[C^{b^{l}_{k}}(R_{z}(\phi^{k}_{l})), C^{b^{l'}_{k'}}(R_{y}(\theta^{k'}_{l'}) )\right] = 0$ whenever $l < l'$, which allows for a full factorization of $T_{0,n}^{0}$ into a sequence of uniformly controlled $y$ rotations, followed by a sequence of uniformly controlled $z$ rotations, as described in \cite{Mottonen2005}.
\subsection{The quantum Fourier transform}

By reversing the order of the $\Lambda$ operators in the pyramidal decomposition, we can interpret the resulting $\psi$-tree operator as effecting a change of basis.  To see this, we construct a special operator product and consider its action on a given basis state $\lvert b^{n}_{k} \rangle$.  By writing $b_{k}^{n} = x_{1}x_{2} \cdots x_{n}$, where $x_{k}$ is the $k$-th digit of the binary number $b_{k}^{n}$, and considering the sequence $b^{1}_{k_{1}} = x_{1}, b^{2}_{k_{2}} = x_{1}x_{2}, \dots, b_{k_{n}}^{n} = b^{n}_{k} $, we can construct an operator product of the form
\begin{equation}
B^{n}_{k} = U_{0}^{0} C^{b^{1}_{k_{1}}}(U^{k_{1}}_{1}) \cdots C^{b^{n}_{k_{n}}}(U^{k_{n}}_{n}).
\end{equation}
The structure of this product corresponds to a particular branch or path within a $\psi$-tree.  Note that the operators appear in the reverse order from the usual pyramidal decomposition described above.  This ensures that the effect of such a sequence will be to produce a separable superposition, qubit by qubit, as one moves from the leaf node to the root along the path.  The corresponding tree operator, which contains all possible such branches, will transform any arbitrary superposition over basis vectors into a superposition over states in the transformed basis. 

Perhaps the best known basis transformation in this context is the quantum Fourier transform.  Before defining the QFT circuit, we must first transform the basis states so that $\lvert b^{n}_{k} \rangle \rightarrow \lvert \overline{b^{n}_{k}} \rangle$, where $ \overline{b^{n}_{k}} = b^{n}_{\overline{k}} = x_{n} x_{n-1} \dots x_{1}$. This corresponds to a sequence of SWAP operations whose net effect is the reversal of the digits of $b^{n}_{k}$.  In what follows, we write the operator sequence in terms of $\overline{k}$, but drop the bar.  

Because the branches $B^{n}_{k}$ each generate separable states, it is straightforward to determine a set of single qubit unitaries $U^{k}_{l}$ in $SU(2)$ whose action produces the desired transformation.    For the QFT, we find that, for interior nodes, the associated unitaries have the form $U^{k}_{l} = (-1)^{k} e^{i\frac{\phi^{k}_{l}}{4}} R_{z}(\frac{\phi^{k}_{l}}{2}) R_{y}(\frac{\pi}{2})$, where $\phi^{k}_{l} = \frac{2 \pi k}{2^{l}}$, while for the leaf nodes, $U_{n}^{k} = (-1)^{k}$.  That is, all unitaries within a given level of the tree exhibit periodicity as a function of $k$.  This periodicity is directly responsible for the polynomial scaling of the number of two-qubit operators in the QFT, as we now show.

First, note that $e^{i\frac{\phi^{k}_{l}}{4}} R_{z}(\frac{\phi^{k}_{l}}{2}) = R(\frac{\phi^{k}_{l}}{2})$, where $R(\phi) =\begin{pmatrix} 1 & 0 \\ 0 & e^{i\phi} \end{pmatrix}$ is the phase shift gate.  Further, observe that we can always factor out an operator of the form $P^{l-1}_{\uparrow} - P^{l-1}_{\downarrow} = Z_{l-1}$ from $\Lambda^{0}_{0,l}$ due to the $(-1)^{k}$ factor in $U^{k}_{l}$, as well as a factor $R^{l}_{y}(\frac{\pi}{2})$ on the right (we insert the superscript to distinguish the qubits they act upon).  

At this stage, it is expedient to define a new operator $\hat{\Lambda}^{j}_{l}$ having the same structure as $\Lambda^{0}_{0,l}$ in terms of the projectors, but containing, in place of the $U^{k}_{l}$ operators, the phase shift gates $R^{j}(\phi^{k}_{l})$, where $j \geq l$ denotes the $j$-th qubit.  This enables us to write $\Lambda^{0}_{0,l} = Z_{l-1} \hat{\Lambda}^{l}_{l} R^{l}_{y}(\frac{\pi}{2})$.   Finally, we note that, because of the periodic nature of the parameters, we can always factor out an operator of the form $(P^{l-1}_{\uparrow} + P^{l-1}_{\downarrow} R(\frac{\Delta \phi_{l}}{2}))$, where $\Delta \phi_{l} = \frac{2 \pi}{2^{l}}$, from $\hat{\Lambda}^{l}_{l}$.  In fact, it is not difficult to confirm that, more generally,
\begin{equation}
\hat{\Lambda}^{k}_{l} =  (P^{l-1}_{\uparrow} + P^{l-1}_{\downarrow} R^{k} (\frac{\Delta \phi_{l}}{2})) \hat{\Lambda}^{k}_{l-1}
\end{equation}
and hence that
\begin{eqnarray}
\Lambda^{0}_{0,l} &=& Z_{l-1} \left( \prod_{k = l}^{1} (P^{k-1}_{\uparrow} + P^{k-1}_{\downarrow} R^{l} (\frac{\Delta \phi_{k}}{2})) \right) R_{y}^{l}(\frac{\pi}{2}) \nonumber \\
&=& Z_{l-1} \prod_{k = l}^{1} C^{k-1}(R^{l} (\frac{\Delta \phi_{k}}{2})) R^{l}_{y}(\frac{\pi}{2}).
\end{eqnarray}
As the full quantum circuit is a product over these levels, and since $R_{y}(\frac{\pi}{2}) Z = H$, where $H$ is the Hadamard operator, we see that the above decomposition matches what appears in the standard treatment of the QFT.  Moreover, it is clear from this derivation that the quadratic scaling in the number of controlled gates with $n$ is a direct consequence of the periodic structure of the rotation parameters appearing in the tree.

\section{Redundant substructure and local changes of basis}

To this point, we have studied how the nested entanglement structure prescribes patterns of gate decompositions for a fixed computational basis.  Here we consider, first, how local (single-qubit) basis transformations affect the state preparation problem in a general sense, and second, how such changes of basis can be used to remove certain redundant substructures from the tree representation of a given state, even in cases where the redundancy is only approximate.  We then discuss the application of a generalized Schmidt decomposition on the largest subtrees that remain.

When $n$ is small, local basis transformations can reduce the overall gate count by a significant margin.  To illustrate, let us first consider the case $n=2$. A Schmidt decomposition of a general two-qubit wavefunction $\lvert \Psi_{2} \rangle$ reduces the corresponding tree from four down to two branches, so that the gate decomposition, by either the subtree or pyramidal method, becomes 
\begin{eqnarray}
\lvert \Psi_{2} \rangle &=& C^{0}(R_{y}^{1}(\pi)) R_{y}^{0} (\theta) \lvert 0 \rangle \nonumber \\
&=&  C^{0}_{1} R_{y}^{0}(\theta) \lvert 0 \rangle.
\end{eqnarray}
Here $\theta = 2\tan^{-1}(\frac{\lambda_{+}}{\lambda_{-}})$, where $\lambda_{\pm}$ are the two eigenvalues of the Schmidt decomposition, and the last line follows from the fact that  $R_{y}(\pi)\cdot Z = X$ and $C^{0}(Z_{1}) \lvert \downarrow \uparrow \rangle = \lvert \downarrow \uparrow \rangle$.  Hence, for two qubits at least, the number of required CNOTs, as compared to either of the two decomposition schemes considered above, can be reduced by half through a judicious choice of basis.

It is possible to generalize the Schmidt decomposition to the case of $n>2$ qubits \cite{Carteret2000}, though the parameters of such a representation are not straightforward to calculate.  Further, while the existence of such representations can be demonstrated through a variational argument, the nonlinear eigenvalue equation derived from this argument admits no general method of solution.  Moreover, such representations can remove at most $3n+1$ parameters from the register wavefunction, a fairly modest reduction when $n$ is large.

One notable exception to these general observations occurs when the tree has redundant substructure; that is, when the magnitude of the overlap between two different subtrees is near to 1.  In such a case, it is possible to ``prune away" one of these branches, at least approximately, by means of a single-qubit basis transformation rendering all nodes within one of these subtrees into dead nodes.  This is most easily seen when the two subtrees constitute adjacent branches, in which case it becomes evident, by equation (\ref{recursive_psi}), that the wavefunction $\lvert \Psi_{n-k} \rangle$ can be factored into a product in the limit $| \langle \Psi_{n-k-1} \lvert \Psi'_{n-k-1} \rangle | \rightarrow 1$. This implies that the corresponding tree reduces to only the leftmost branch under a trivial basis transformation of the first qubit.  

More generally, when the two subtrees are not adjacent, one may still prune one of them away by means of a local change of basis, provided the paths to each subtree differ by only one spin flip.  At the qubit where the two paths diverge, one may choose a basis such that the path to one of the two subtrees has an amplitude of zero.  This can be seen by focusing only on that part of the register wavefunction $\lvert \Psi_{n} \rangle$ corresponding to the two paths containing the subtrees $\vert \Psi_{1} \rangle$ and $\lvert \Psi_{2} \rangle$, i.e.,
\begin{eqnarray}
\lvert \Psi_{1,2} \rangle &=& \alpha \lvert \cdots \uparrow \cdots \rangle \lvert \Psi_{1} \rangle + \beta \lvert \cdots \downarrow \cdots \rangle \lvert \Psi_{2} \rangle
\end{eqnarray}
where $\alpha, \beta$ are the (relative) amplitudes of the two branches, normalized so that $|\alpha|^{2} + |\beta|^{2} = 1$, and where the states $\vert \cdots \uparrow \cdots \rangle, \lvert \cdots \downarrow \cdots \rangle$ are identical except for the indicated qubit.  If we define $\kappa = \langle \Psi_{1} | \Psi_{2} \rangle$, then, in the limit that $|\kappa|$ is close to one, the Schmidt decomposition of $\lvert \Psi_{1,2} \rangle$ becomes 
\begin{equation}
\lvert \Psi_{1,2} \rangle \rightarrow \lambda_{+} \lvert \cdots \widehat{\uparrow} \cdots \rangle \lvert \Psi_{1} \rangle + \lambda_{-} \lvert \cdots \widehat{\downarrow} \cdots \rangle \frac{(\lvert \Psi_{2} \rangle  - \kappa \lvert \Psi_{1} \rangle)}{\sqrt{1 - |\kappa|^{2}}
}\end{equation}
where  $\lambda_{\pm}$ are the (real-valued) roots of
\begin{equation}
\lambda^{2} - \lambda + |\alpha|^{2} |\beta|^{2}(1 - |\kappa|^{2}) = 0,
\end{equation}
and the hat indicates that the corresponding qubits are being represented in the new basis.  In the limit that $|\kappa| \rightarrow 1$, $\lambda_{+} \rightarrow 1$ and $\lambda_{-} \rightarrow 0$, and we may safely ignore the second branch of the wavefunction.  When $|\kappa|$ is near to but less than 1, this provides a serviceable approximation to the wavefunction and allows for a more parsimonious gate decomposition at the cost of reduced fidelity. 

When two redundant, or nearly redundant, subtrees lie on paths differing by more than one spin flip, it is always possible to rearrange the branches of the tree so that this condition becomes satisfied.  Such rearrangements can be accomplished by factoring out an appropriate number of multiply-controlled CNOTs, which have the effect of trading the two branches at the node specified by the control sequence.  Whether such a move is advantageous depends on the balance of entangling gates created by the rearrangement process, to those removed by pruning.  

Finally, after removing any redundant substructure from the tree, one may proceed to perform a generalized Schmidt decomposition on any of the largest remaining subtrees still in the original basis.  As discussed by Carteret et. al. \cite{Carteret2000} using a variational argument, such a transformation zeroes out all coefficients of the basis states differing from $\lvert 0 \rangle$ by a single spin flip, and eliminates the phases from the coefficients of their time-reversed partners.  If we take $m$ to be the number of qubits composing the subtree wavefunction $\lvert \Omega_{m} \rangle = \sum_{k=0}^{M-1} \gamma_{k} \lvert b^{m}_{k} \rangle$, where $M = 2^{m}$, this removes $3m+ 1$ parameters.  In what follows, we depart from the usual variational approach, which leads to a nonlinear eigenvalue equation for the coefficients, and instead derive a system of simultaneous multilinear equations for the transformation parameters themselves.

Within this subtree, we consider a combination of single-qubit basis transformations of the form
\begin{eqnarray}
\lvert \widehat{\uparrow} \rangle_{i} &=& \alpha_{i} \lvert \uparrow \rangle_{i} + \beta_{i} \lvert \downarrow \rangle_{i} \nonumber \\
\lvert \widehat{\downarrow} \rangle_{i} &=& \alpha_{i}^{*} \lvert \downarrow \rangle_{i} - \beta_{i}^{*} \lvert \uparrow \rangle_{i}
\end{eqnarray}
for all qubits $i$ in the subtree register, where the coefficients $\alpha_{i}, \beta_{i}$ are parameterized as
\begin{eqnarray}
\alpha_{i} &=& \cos (\frac{\theta_{i}}{2}) e^{\frac{i}{2} (\phi_{i} - \chi_{i})} \nonumber \\
\beta_{i} &=& \sin (\frac{\theta_{i}}{2}) e^{\frac{i}{2} (\phi_{i} + \chi_{i})}.
\end{eqnarray}
In the first stage of the calculation, we focus on the subset of states $\lvert b^{m}_{2^{k}} \rangle$, where $k$ ranges from $0$ to $m-1$.  These define a subspace of states differing from $\lvert 0 \rangle$ by single spin flip.  We require the $m$ amplitudes $\widehat{\gamma}_{2^{k}}$ in the new basis, which can be found by computing
\begin{eqnarray}
\widehat{\gamma}_{2^{k}} &=& \langle \widehat{b}^{m}_{2^{k}} |\Omega_{m} \rangle \nonumber \\
&=& \sum_{i} \gamma_{i} \langle \widehat{b}^{m}_{2^{k}} | b^{m}_{i} \rangle 
\end{eqnarray}
where the amplitudes $ \langle \widehat{b}^{m}_{2^{k}} | b^{m}_{i} \rangle $ are given by 
\begin{eqnarray}
\langle \widehat{b}^{m}_{2^{k}} | b^{m}_{i} \rangle &=& \langle \widehat{b}^{m}_{2^{k}} | x_{0} x_{1} \cdots x_{m-1} \rangle \nonumber \\
&=& f_{k}' \prod_{j \neq k} f_{j}.
\end{eqnarray}
Here, as above, $x_{j}$ is the $j$-th digit in the binary representation of $b_{i}^{m}$, and the $f_{i}$ are given by
\begin{eqnarray}
f_{i} &=& \alpha_{i} - x_{i} (\beta^{*}_{i} + \alpha_{i}) \nonumber \\
f_{i}' &=& \beta_{i} + x_{i}(\alpha_{i}^{*} - \beta_{i}).
\end{eqnarray}

Setting $\widehat{\gamma}_{2^{k}}=0$ and dividing through by $\alpha_{k}^{*} \prod_{i \neq k}^{m-1} \alpha_{i}$ yields an equation of the form
\begin{equation}
0 = \gamma^{(0)}_{0} +  \sum_{i=0}^{m-1} \gamma^{(1)}_{i} z_{i} + \sum_{i<j}^{m-1} \gamma_{i,j}^{(2)} z_{i} z_{j} + \cdots
\end{equation}
where $z_{i} =  \frac{\beta_{i}}{\alpha_{i}^{*}}$ if $i = k$, and  $- \frac{\beta_{i}^{*}}{\alpha_{i}}$ otherwise, and we have relabeled the $\gamma$ coefficients so that the superscript refers to the number of spin flips separating the associated basis vector from $\lvert b^{m}_{2^{k}} \rangle$, and the subscripts label the specific qubits where they differ.  As there are $m$ such equations in the $m$ complex numbers $\tan \frac{\theta_{i}}{2} e^{i \phi_{i}}$ (and their complex conjugates), these constitute a complex multilinear system for determining the parameters $\theta_{i}, \phi_{i}$.  The parameters $\chi_{i}$ can then be chosen to eliminate the phases of the coefficients of the time-reversed partners of the   $\lvert b^{m}_{2^{k}} \rangle$.

\section{Summary and Discussion}
In this paper, the state of a quantum register is represented in terms of the nested entanglement of each qubit with its successors in an ordered list.  This approach leads naturally to a binary tree construction whose parameterization captures the entanglement structure of the wavefunction in a parsimonious way.  The parameters of this tree are calculated by means of a simple algorithm that converts the coefficients of an arbitrary wavefunction in the standard computational basis into the complex weights of the nested representation.  The resulting data structure can be used to automate the construction of circuits that prepare the associated quantum state starting from $\lvert 0 \rangle$.

We have explored two distinct circuit decomposition schemes arising from the recursive substructure of the tree.  In both cases, gate sequences are constructed by means of recursion relations and elementary operator identities.  The ``subtree" decomposition leads to a simple separability criterion for arbitrary pure states when their corresponding $\psi$-trees are represented in a particular canonical form.  The ``pyramidal" decomposition is shown to be equivalent to the formalism of uniformly controlled rotations developed by M{\"o}tt{\"o}nen et. al.  Elements of this latter decomposition are also used to construct the standard circuit for the quantum Fourier transform.

Finally, we discussed local basis transformations of the qubits and the removal of redundant substructure from $\psi$-trees, including methods for approximating the corresponding wavefunction when two subtrees strongly overlap.  We also discussed the generalized Schmidt decomposition as a method for removing additional parameters from a subtree lacking redundant substructure.

Based on the foregoing investigation of the nested entanglement formalism, especially the discussions surrounding the construction of the quantum Fourier transform and the removal of redundant substructure, there would seem to be a correspondence between the  ``compressibility" of a given wavefunction's $\psi$-tree description and the number of entangling gates required to produce it.  In the case of the quantum Fourier transform, this manifests as the breakdown of its pyramidal decomposition into a polynomial number of entangling factors.  In the case of a general wavefunction, it can take the form of repeated subtrees, periodicities, or other ordered patterns of tree parameters.  

As the $\psi$-tree representation of most random wavefunctions will be incompressible, they can be of little practical use in quantum computing applications for large $n$, since the quantum circuits required to produce them will require $\sim 2^{n}$ entangling gates.  This suggests that the state preparation subroutine will be restricted to wavefunctions with an ordered tree representation, featuring periodicities, repeated subtrees, dead branches, or other such structure to achieve reasonable scaling.  The nature and multiplicity of the structures required to ensure polynomial scaling is an interesting question for future work.

\begin{acknowledgments}
The author gratefully acknowledges helpful comments from Richard Preston, Yaakov Weinstein, and Gerry Gilbert.  This work was funded by MITRE's Independent Research and Development Program.  

\textbf{Approved for Public Release; Distribution Unlimited.  Public Release Case Number 24-0359.}  \textcopyright \textbf{2024 The MITRE Corporation.  ALL RIGHTS RESERVED.}
\end{acknowledgments}

\bibliography{quantum_control}

\providecommand{\noopsort}[1]{}\providecommand{\singleletter}[1]{#1}%
\begin{thebibliography}{15}%
\makeatletter
\providecommand \@ifxundefined [1]{%
 \@ifx{#1\undefined}
}%
\providecommand \@ifnum [1]{%
 \ifnum #1\expandafter \@firstoftwo
 \else \expandafter \@secondoftwo
 \fi
}%
\providecommand \@ifx [1]{%
 \ifx #1\expandafter \@firstoftwo
 \else \expandafter \@secondoftwo
 \fi
}%
\providecommand \natexlab [1]{#1}%
\providecommand \enquote  [1]{``#1''}%
\providecommand \bibnamefont  [1]{#1}%
\providecommand \bibfnamefont [1]{#1}%
\providecommand \citenamefont [1]{#1}%
\providecommand \href@noop [0]{\@secondoftwo}%
\providecommand \href [0]{\begingroup \@sanitize@url \@href}%
\providecommand \@href[1]{\@@startlink{#1}\@@href}%
\providecommand \@@href[1]{\endgroup#1\@@endlink}%
\providecommand \@sanitize@url [0]{\catcode `\\12\catcode `\$12\catcode
  `\&12\catcode `\#12\catcode `\^12\catcode `\_12\catcode `\%12\relax}%
\providecommand \@@startlink[1]{}%
\providecommand \@@endlink[0]{}%
\providecommand \url  [0]{\begingroup\@sanitize@url \@url }%
\providecommand \@url [1]{\endgroup\@href {#1}{\urlprefix }}%
\providecommand \urlprefix  [0]{URL }%
\providecommand \Eprint [0]{\href }%
\providecommand \doibase [0]{https://doi.org/}%
\providecommand \selectlanguage [0]{\@gobble}%
\providecommand \bibinfo  [0]{\@secondoftwo}%
\providecommand \bibfield  [0]{\@secondoftwo}%
\providecommand \translation [1]{[#1]}%
\providecommand \BibitemOpen [0]{}%
\providecommand \bibitemStop [0]{}%
\providecommand \bibitemNoStop [0]{.\EOS\space}%
\providecommand \EOS [0]{\spacefactor3000\relax}%
\providecommand \BibitemShut  [1]{\csname bibitem#1\endcsname}%
\let\auto@bib@innerbib\@empty
\bibitem [{\citenamefont {Vidal}(2003)}]{Vidal2003}%
  \BibitemOpen
  \bibfield  {author} {\bibinfo {author} {\bibfnamefont {G.}~\bibnamefont
  {Vidal}},\ }\bibfield  {title} {\bibinfo {title} {Efficient classical
  simulation of slightly entangled quantum computations},\ }\href@noop {}
  {\bibfield  {journal} {\bibinfo  {journal} {Phys. Rev. Lett.}\ }\textbf
  {\bibinfo {volume} {91}},\ \bibinfo {pages} {147902} (\bibinfo {year}
  {2003})}\BibitemShut {NoStop}%
\bibitem [{\citenamefont {Jozsa}\ and\ \citenamefont
  {Linden}(2003)}]{Jozsa2003}%
  \BibitemOpen
  \bibfield  {author} {\bibinfo {author} {\bibfnamefont {R.}~\bibnamefont
  {Jozsa}}\ and\ \bibinfo {author} {\bibfnamefont {N.}~\bibnamefont {Linden}},\
  }\bibfield  {title} {\bibinfo {title} {On the role of entanglement in
  quantum-computational speed-up},\ }\href@noop {} {\bibfield  {journal}
  {\bibinfo  {journal} {Proc. R. Soc. Lond. Ser. A: Math. Phys. Eng.}\ }\textbf
  {\bibinfo {volume} {459}} (\bibinfo {year} {2003})}\BibitemShut {NoStop}%
\bibitem [{\citenamefont {Plesch}\ and\ \citenamefont
  {Brukner}(2011)}]{Plesch2011}%
  \BibitemOpen
  \bibfield  {author} {\bibinfo {author} {\bibfnamefont {M.}~\bibnamefont
  {Plesch}}\ and\ \bibinfo {author} {\bibfnamefont {C.}~\bibnamefont
  {Brukner}},\ }\bibfield  {title} {\bibinfo {title} {Quantum-state preparation
  with universal gate decompositions},\ }\href@noop {} {\bibfield  {journal}
  {\bibinfo  {journal} {Phys.\ Rev. A}\ }\textbf {\bibinfo {volume} {83}},\
  \bibinfo {pages} {032302} (\bibinfo {year} {2011})}\BibitemShut {NoStop}%
\bibitem [{\citenamefont {Znidaric}\ \emph {et~al.}(2008)\citenamefont
  {Znidaric}, \citenamefont {Giraud},\ and\ \citenamefont
  {Georgeot}}]{Znidaric2008}%
  \BibitemOpen
  \bibfield  {author} {\bibinfo {author} {\bibfnamefont {M.}~\bibnamefont
  {Znidaric}}, \bibinfo {author} {\bibfnamefont {O.}~\bibnamefont {Giraud}},\
  and\ \bibinfo {author} {\bibfnamefont {B.}~\bibnamefont {Georgeot}},\
  }\bibfield  {title} {\bibinfo {title} {Optimal number of controlled-{NOT}
  gates to generate a three-qubit state},\ }\href@noop {} {\bibfield  {journal}
  {\bibinfo  {journal} {Phys. \ Rev. A}\ }\textbf {\bibinfo {volume} {77}},\
  \bibinfo {pages} {042320} (\bibinfo {year} {2008})}\BibitemShut {NoStop}%
\bibitem [{\citenamefont {Kaye}\ and\ \citenamefont {Mosca}()}]{Kaye2001}%
  \BibitemOpen
  \bibfield  {author} {\bibinfo {author} {\bibfnamefont {P.}~\bibnamefont
  {Kaye}}\ and\ \bibinfo {author} {\bibfnamefont {M.}~\bibnamefont {Mosca}},\
  }\bibfield  {title} {\bibinfo {title} {Quantum networks for generating
  arbitrary quantum states},\ }\href@noop {} {\bibinfo  {journal} {Proceedings
  of the International Conference on Quantum Information (Rochester, New York,
  2001)}\ }\BibitemShut {NoStop}%
\bibitem [{\citenamefont {de~Brougiere}\ \emph {et~al.}(2020)\citenamefont
  {de~Brougiere}, \citenamefont {Baboulin}, \citenamefont {Valiron},\ and\
  \citenamefont {Allouche}}]{Goubault2020}%
  \BibitemOpen
\bibfield  {journal} {  }\bibfield  {author} {\bibinfo {author} {\bibfnamefont
  {T.~G.}\ \bibnamefont {de~Brougiere}}, \bibinfo {author} {\bibfnamefont
  {M.}~\bibnamefont {Baboulin}}, \bibinfo {author} {\bibfnamefont
  {B.}~\bibnamefont {Valiron}},\ and\ \bibinfo {author} {\bibfnamefont
  {C.}~\bibnamefont {Allouche}},\ }\bibfield  {title} {\bibinfo {title}
  {Quantum circuit synthesis using {H}ouseholder transformations},\ }\href@noop
  {} {\bibfield  {journal} {\bibinfo  {journal} {Computer Physics
  Communications}\ }\textbf {\bibinfo {volume} {248}},\ \bibinfo {pages}
  {107001} (\bibinfo {year} {2020})}\BibitemShut {NoStop}%
\bibitem [{\citenamefont {M{\"o}tt{\"o}nen}\ \emph {et~al.}(2004)\citenamefont
  {M{\"o}tt{\"o}nen}, \citenamefont {Vartiainen}, \citenamefont {Bergholm},\
  and\ \citenamefont {Salomaa}}]{Mottonen2004}%
  \BibitemOpen
  \bibfield  {author} {\bibinfo {author} {\bibfnamefont {M.}~\bibnamefont
  {M{\"o}tt{\"o}nen}}, \bibinfo {author} {\bibfnamefont {J.}~\bibnamefont
  {Vartiainen}}, \bibinfo {author} {\bibfnamefont {V.}~\bibnamefont
  {Bergholm}},\ and\ \bibinfo {author} {\bibfnamefont {M.~M.}\ \bibnamefont
  {Salomaa}},\ }\bibfield  {title} {\bibinfo {title} {Quantum circuits for
  general multiqubit gates},\ }\href@noop {} {\bibfield  {journal} {\bibinfo
  {journal} {Phys. Rev. Lett.}\ }\textbf {\bibinfo {volume} {93}},\ \bibinfo
  {pages} {130502} (\bibinfo {year} {2004})}\BibitemShut {NoStop}%
\bibitem [{\citenamefont {M{\"o}tt{\"o}nen}\ \emph {et~al.}(2005)\citenamefont
  {M{\"o}tt{\"o}nen}, \citenamefont {Vartiainen}, \citenamefont {Bergholm},\
  and\ \citenamefont {Salomaa}}]{Mottonen2005}%
  \BibitemOpen
  \bibfield  {author} {\bibinfo {author} {\bibfnamefont {M.}~\bibnamefont
  {M{\"o}tt{\"o}nen}}, \bibinfo {author} {\bibfnamefont {J.}~\bibnamefont
  {Vartiainen}}, \bibinfo {author} {\bibfnamefont {V.}~\bibnamefont
  {Bergholm}},\ and\ \bibinfo {author} {\bibfnamefont {M.~M.}\ \bibnamefont
  {Salomaa}},\ }\bibfield  {title} {\bibinfo {title} {Transformation of quantum
  states using uniformly controlled rotations},\ }\href@noop {} {\bibfield
  {journal} {\bibinfo  {journal} {Quantum Information \& Computation}\ }\textbf
  {\bibinfo {volume} {5}},\ \bibinfo {pages} {467} (\bibinfo {year}
  {2005})}\BibitemShut {NoStop}%
\bibitem [{\citenamefont {Rakyta}\ and\ \citenamefont
  {Zimbor{\'a}s}(2022)}]{Rakyta2022}%
  \BibitemOpen
  \bibfield  {author} {\bibinfo {author} {\bibfnamefont {P.}~\bibnamefont
  {Rakyta}}\ and\ \bibinfo {author} {\bibfnamefont {Z.}~\bibnamefont
  {Zimbor{\'a}s}},\ }\bibfield  {title} {\bibinfo {title} {Approaching the
  theoretical limit in quantum gate decomposition},\ }\href@noop {} {\bibfield
  {journal} {\bibinfo  {journal} {Quantum}\ }\textbf {\bibinfo {volume} {6}},\
  \bibinfo {pages} {710} (\bibinfo {year} {2022})}\BibitemShut {NoStop}%
\bibitem [{\citenamefont {Carteret}\ \emph {et~al.}(2000)\citenamefont
  {Carteret}, \citenamefont {Higuchi},\ and\ \citenamefont
  {Sudbery}}]{Carteret2000}%
  \BibitemOpen
  \bibfield  {author} {\bibinfo {author} {\bibfnamefont {H.~A.}\ \bibnamefont
  {Carteret}}, \bibinfo {author} {\bibfnamefont {A.}~\bibnamefont {Higuchi}},\
  and\ \bibinfo {author} {\bibfnamefont {A.}~\bibnamefont {Sudbery}},\
  }\bibfield  {title} {\bibinfo {title} {Multipartite generalization of the
  {S}chmidt decomposition},\ }\href@noop {} {\bibfield  {journal} {\bibinfo
  {journal} {Journal of Mathematical Physics}\ }\textbf {\bibinfo {volume}
  {41}},\ \bibinfo {pages} {7932} (\bibinfo {year} {2000})}\BibitemShut
  {NoStop}%
\bibitem [{\citenamefont {Jorrand}\ and\ \citenamefont
  {Mhalla}(2003)}]{Jorrand2003}%
  \BibitemOpen
  \bibfield  {author} {\bibinfo {author} {\bibfnamefont {P.}~\bibnamefont
  {Jorrand}}\ and\ \bibinfo {author} {\bibfnamefont {M.}~\bibnamefont
  {Mhalla}},\ }\bibfield  {title} {\bibinfo {title} {Separability of pure
  n-qubit states: two characterizations},\ }\href@noop {} {\bibfield  {journal}
  {\bibinfo  {journal} {Int. J. Found. Comput. Sci.}\ }\textbf {\bibinfo
  {volume} {14}},\ \bibinfo {pages} {797} (\bibinfo {year} {2003})}\BibitemShut
  {NoStop}%
\bibitem [{\citenamefont {shui Yu}\ and\ \citenamefont {shan
  Song}(2006)}]{Yu2006}%
  \BibitemOpen
  \bibfield  {author} {\bibinfo {author} {\bibfnamefont {C.}~\bibnamefont {shui
  Yu}}\ and\ \bibinfo {author} {\bibfnamefont {H.}~\bibnamefont {shan Song}},\
  }\bibfield  {title} {\bibinfo {title} {Global entanglement for multipartite
  quantum states},\ }\href@noop {} {\bibfield  {journal} {\bibinfo  {journal}
  {Phys. Rev. A}\ }\textbf {\bibinfo {volume} {73}},\ \bibinfo {pages} {022325}
  (\bibinfo {year} {2006})}\BibitemShut {NoStop}%
\bibitem [{\citenamefont {M{\"a}kel{\"a}}\ and\ \citenamefont
  {Messina}(2010)}]{Makela2010}%
  \BibitemOpen
  \bibfield  {author} {\bibinfo {author} {\bibfnamefont {H.}~\bibnamefont
  {M{\"a}kel{\"a}}}\ and\ \bibinfo {author} {\bibfnamefont {A.}~\bibnamefont
  {Messina}},\ }\bibfield  {title} {\bibinfo {title} {Polynomial method to
  study the entanglement of pure {N}-qubit states},\ }\href@noop {} {\bibfield
  {journal} {\bibinfo  {journal} {Phys. Rev. A}\ }\textbf {\bibinfo {volume}
  {81}},\ \bibinfo {pages} {012326} (\bibinfo {year} {2010})}\BibitemShut
  {NoStop}%
\bibitem [{\citenamefont {Neven}\ and\ \citenamefont
  {Bastin}(2018)}]{Neven2018}%
  \BibitemOpen
  \bibfield  {author} {\bibinfo {author} {\bibfnamefont {A.}~\bibnamefont
  {Neven}}\ and\ \bibinfo {author} {\bibfnamefont {T.}~\bibnamefont {Bastin}},\
  }\bibfield  {title} {\bibinfo {title} {The quantum separability problem is a
  simultaneous hollowisation matrix analysis problem},\ }\href@noop {}
  {\bibfield  {journal} {\bibinfo  {journal} {J. Phys. A: Math. Theor.}\
  }\textbf {\bibinfo {volume} {51}},\ \bibinfo {pages} {315305} (\bibinfo
  {year} {2018})}\BibitemShut {NoStop}%
\bibitem [{\citenamefont {Horodecki}\ \emph {et~al.}(2009)\citenamefont
  {Horodecki}, \citenamefont {Horodecki}, \citenamefont {Horodecki},\ and\
  \citenamefont {Horodecki}}]{Horodecki2009}%
  \BibitemOpen
  \bibfield  {author} {\bibinfo {author} {\bibfnamefont {R.}~\bibnamefont
  {Horodecki}}, \bibinfo {author} {\bibfnamefont {P.}~\bibnamefont
  {Horodecki}}, \bibinfo {author} {\bibfnamefont {M.}~\bibnamefont
  {Horodecki}},\ and\ \bibinfo {author} {\bibfnamefont {K.}~\bibnamefont
  {Horodecki}},\ }\bibfield  {title} {\bibinfo {title} {Quantum entanglement},\
  }\href@noop {} {\bibfield  {journal} {\bibinfo  {journal} {Rev. Mod. Phys.}\
  }\textbf {\bibinfo {volume} {81}} (\bibinfo {year} {2009})}\BibitemShut
  {NoStop}%
\end{thebibliography}%

\end{document}